\documentclass[float,epsfig,usenatbib]{mn2e}

\usepackage{graphicx}
\usepackage{appendix}
\usepackage{psfrag}
\usepackage{amsmath,amssymb}
\usepackage{threeparttable}
\usepackage{float}
\usepackage{subfigure}
\usepackage{afterpage}

\def\apj{ApJ}
\def\apjl{ApJL}
\def\mnras{MNRAS}
\def\pasp{PASP}
\def\aj{AJ}
\def\araa{ARA\&A}
\def\aap{A\&A}
\def\apjs{ApJS}

\title[The effect of the environment on the HI scaling relations]{The effect of the environment on the \hi~scaling relations}
\author[L. Cortese et al.]
{L. Cortese\thanks{lcortese@eso.org}$^1$, B. Catinella$^2$, S. Boissier$^3$, A. Boselli$^3$, S. Heinis$^3$\\ 
$^1$European Southern Observatory, Karl-Schwarzschild Str. 2, 85748 Garching bei Muenchen, Germany\\
$^2$Max-Planck Institut fur Astrophysik, D-85741 Garching bei Muenchen, Germany\\
$^3$Laboratoire d'Astrophysique de Marseille, UMR6110 CNRS, 38 rue F. Joliot-Curie, 13388 Marseille, France}

\date{}

\begin{document}
\date{Accepted 2011 March 29.  Received 2011 March 28; in original form 2011 March 9}
\newcommand{\Zsolar}{\mbox{$\,\rm Z_{\odot}$}}
\newcommand{\Msolar}{\mbox{$\,\rm M_{\odot}$}}
\newcommand{\Lsolar}{\mbox{$\,\rm L_{\odot}$}}
\newcommand{\xs}{$\chi^{2}$}
\newcommand{\dxs}{$\Delta\chi^{2}$}
\newcommand{\xsn}{$\chi^{2}_{\nu}$}
\newcommand{\ls}{{\tiny \( \stackrel{<}{\sim}\)}}
\newcommand{\gs}{{\tiny \( \stackrel{>}{\sim}\)}}
\newcommand{\asec}{$^{\prime\prime}$}
\newcommand{\amin}{$^{\prime}$}
\newcommand{\mstar}{\mbox{$M_{*}$}}
\newcommand{\hi}{H{\sc i}}
\newcommand{\hii}{H{\sc ii}\ }
\newcommand{\kms}{km~s$^{-1}$\ }

\maketitle

\label{firstpage}

\begin{abstract}
We use a volume-, magnitude-limited sample of nearby galaxies to investigate the effect of the environment 
on the \hi~scaling relations. 
We confirm that the \hi-to-stellar mass ratio anti correlates with stellar mass, 
stellar mass surface density and $NUV-r$ colour across the whole range of parameters 
covered by our sample (10$^{9}$ \ls$M_{*}$\ls 10$^{11}$ M$_{\odot}$, 7.5 \ls$\mu_{*}$\ls 9.5 M$_{\odot}$ kpc$^{-2}$, 2 \ls$NUV-r$\ls 6 mag). 
These scaling relations are also followed by galaxies in the Virgo cluster, although they are significantly offset 
towards lower gas content. 
Interestingly, the difference between field and cluster galaxies gradually decreases moving towards massive, bulge-dominated 
systems. By comparing our data with the predictions of chemo-spectrophotometric models of galaxy 
evolution, we show that starvation alone cannot explain the low gas content of Virgo 
spirals and that only ram-pressure stripping is able to reproduce our findings.
Finally, motivated by previous studies, we investigate the use of a plane obtained from the relations 
between the \hi-to-stellar mass ratio, stellar mass surface density and $NUV-r$ colour as a proxy for 
the \hi~deficiency parameter. We show that the distance from the `\hi~gas fraction plane' can be used 
as an alternative estimate for the \hi~deficiency, but only if carefully calibrated on pre-defined samples 
of `unperturbed' systems. 
\end{abstract}


\section{Introduction}
Understanding the role played by the cold gas component in 
galaxy evolution is one of the main challenges for 
extragalactic studies. The cold atomic hydrogen (\hi) 
is the reservoir out of which all stars are eventually formed.
Therefore a detailed knowledge of its relation to other galaxy properties 
is of primary importance in order to build a coherent picture 
of galaxy evolution.

Although the direct detection of \hi~emission from individual galaxies is still 
technically limited to the nearby universe (e.g., \citealp{verhaijen07,catinella08b}), in the last decades several 
studies have highlighted how the \hi~content (usually estimated 
as the \hi~mass-to-luminosity ratio) varies with 
morphology, luminosity, size and star formation activity 
(e.g., \citealp{roberts63,haynes,roberts94,phenomen,kannappan04,mcgaugh97,boselli01}).
Particularly powerful appears to be the use of 21 cm data to 
investigate environmental effects. Several studies have shown 
that \hi-deficient\footnote{The 
H{\sc i}-deficiency ($Def_{HI}$) is defined as the difference, in 
logarithmic units, between the expected H{\sc i} mass for an isolated galaxy 
with the same morphological type and optical diameter of the target and the observed value \citep{haynes}.} 
galaxies are mainly/only present in very high-density environments \citep{giova85,cayatte90,solanes01,bosellicomaI}, 
providing strong constraints on the possible environmental mechanisms 
responsible for the quenching of the star formation in clusters 
of galaxies (see \citealp{review} and references therein).
 
Unfortunately, a great part of what we know about \hi~in galaxies has been achieved 
by studying relatively small samples, with not always well defined selection criteria, 
often focused on specific morphological classes (e.g., only late-type galaxies) 
and with small overlap with multiwavelength surveys, necessary to investigate 
how the cold gas correlates with other baryonic components.
Moreover, it is very difficult for simulations to provide accurate estimates 
for important observables like the \hi~deficiency parameter (since it is based on a 
detailed morphological classification), limiting the comparison between observations and theory. 
In other words, \hi~astronomy is in a situation similar to the 
one of optical and ultraviolet astronomy before the Sloan Digital Sky Survey 
and the GALEX mission. The main scaling relations (e.g., colour-magnitude, 
size-luminosity, etc.) were already known, but a detailed quantification 
of their properties, crucial for a comparison with models, was still missing. 

Luckily, the situation is rapidly changing for 21 cm studies. 
The advent of large \hi~surveys, such as the {\it Arecibo Legacy Fast ALFA Survey} \citep{alfaalfa05}, 
which eventually will provide \hi~masses for a few tens of thousands of galaxies over areas of sky with large multiwavelength coverage, 
is gradually allowing detailed statistical analysis of \hi~properties in the local 
universe. 
Very promising results in this direction have been recently obtained 
by the {\it GALEX Arecibo SDSS Survey} (GASS; \citealp{catinella10}), a targeted survey 
of a volume-limited sample of $\sim$1000 massive galaxies ($M_{*}>$10$^{10}$ M$_{\odot}$).
\cite{catinella10} used the first GASS data release to quantify the main scaling relations 
linking the \hi-to-stellar mass ratio to stellar mass, stellar mass surface density and colour. 
They also suggested the existence of a `gas fraction plane' linking \hi~gas 
fraction\footnote{In this paper we will refer to the $M(HI)/M_{*}$ ratio as gas fraction.} 
to stellar mass surface density and $NUV-r$ colour and proposed to use this plane 
to isolate interesting outliers that might be in the process of accreting or loosing a significant 
fraction of their gas content. In this context, the plane might be considered as an alternative 
way to define \hi~deficiency when accurate morphological classification is not available. 
However, this hypothesis has never been tested for a sample for which both \hi~deficiency 
and the \hi~gas fraction plane can be estimated independently. 

In this paper, we use a volume-, magnitude-limited sample of $\sim$300 galaxies to 
extend the study of the \hi~scaling relations to a larger dynamical range in stellar mass 
and stellar mass surface density and to different environments (i.e., from isolated systems 
to the center of the Virgo cluster). 
Our main goals are a) to accurately quantify the effect of the cluster environment on the 
\hi~scaling relations, b) to investigate the validity of the \hi~plane as 
a proxy for \hi~deficiency and c) to highlight the power of \hi~scaling relations 
to constrain models of galaxy formation and evolution.  

\section{The sample}
The analysis presented in this paper is based on the Herschel Reference Survey (HRS, \citealp{HRS}).
Briefly, this consists of a volume-limited sample (i.e., 15$\leq d \leq$25 Mpc) including late-type galaxies (Sa and later) 
with  2MASS \citep{2massall} K-band magnitude K$_{Stot} \le$ 12 mag and early-type galaxies (S0a and earlier) with 
K$_{Stot} \le$ 8.7 mag. Additional selection criteria are high galactic latitude (b $>$ +55$^{\circ}$) and low Galactic extinction (A$_{B}$ $<$ 0.2 mag, \citealp{schlegel98}), 
to minimize Galactic cirrus contamination.
The total sample consists of 322 galaxies (260 late- and 62 early-type galaxies) \footnote{HRS228/LCRS B123647.4-052325 has been removed from the original sample since 
the redshift reported in NED was not correct and the galaxy lies well beyond the limits of the HRS, $z\sim$0.04.}.
As extensively discussed in \cite{HRS}, this sample is not only representative of the local universe but, spanning 
different density regimes (i.e., from isolated systems to the center of the Virgo cluster), it is also ideal for environmental 
studies (see also \citealp{cortese09}).

Atomic hydrogen masses have been estimated from \hi~21 cm line emission data (mainly single-dish), available from the 
literature (e.g., \citealp{spring05hi,giovanelli07,kent08,goldmine} and the {\it NASA/IPAC Extragalactic Database}, NED).
In total, 305 out of 322 galaxies in the HRS ($\sim$95\%) have been observed at 21 cm, with 265 detections and 40 non detections.  
\hi~masses have been computed via
\begin{equation}
\frac{M(HI)}{M_{\odot}}= 2.356 \times 10^{5} \frac{S_{HI}}{\rm Jy~km s^{-1}} \Big(\frac{d}{\rm Mpc}\Big)^{2} 
\end{equation}
where $S_{HI}$ is the integrated \hi~line flux-density and $d$ is the distance. 
As discussed in \cite{HRS}, we fixed the distances for galaxies belonging to the Virgo cluster (i.e., 23 Mpc for the 
Virgo B cloud and 17 Mpc for all the other clouds; \citealp{gav99}), while for the rest of the sample 
distances have been estimated from their recessional velocities assuming a Hubble constant $H_{0}=$70 km s$^{-1}$ Mpc$^{-1}$.
In case of non detections, upper limits have been determined assuming a 5$\sigma$ signal with 300 km s$^{-1}$ velocity 
width. 
We estimate the \hi~deficiency parameter ($Def_{HI}$) following \cite{haynes}.
The expected \hi~mass for each galaxy is determined via 
\begin{equation}
\log(M(HI)_{exp}) = a_{HI} + b_{HI}\times\log\Big(\frac{h D_{25}}{\rm kpc}\Big) -2\log (h) 
\end{equation} 
where $h=H_{0}/100$ km s$^{-1}$ Mpc$^{-1}$, $D_{25}$ is the optical isophotal diameter measured at 25 mag arcsec$^{2}$ in B-band and 
$a_{HI}$ and $b_{HI}$ are two coefficients that vary with morphological type. 
These coefficients have been calculated for the following types (see also Table 3 in \citealp{bosellicomaI}): S0a and earlier 
\citep{haynes}, Sa-Sab, Sb, Sbc, Sc \citep{solanes96} and Scd to Irr \citep{bosellicomaI}.
Morphological classifications for our sample are taken from the Virgo Cluster Catalogue \citep{vcc}, NED and \cite{HRS}.
The \hi~deficiency is then 
\begin{equation}
 Def_{HI}=\log(M(HI)_{exp}) - \log(M(HI))
\end{equation}
We note that the estimate of the expected \hi~mass for early-type galaxies is extremely uncertain. 
As we will show later, this uncertainty can significantly affect the quantification of 
the \hi~scaling relations.
In the following, we will consider as `\hi-deficient' galaxies those objects with $Def_{HI}\geq0.5$ (i.e., 
galaxies with 70\% less hydrogen than isolated systems with the same diameter and morphological type). 
The average \hi~deficiency of galaxies with $Def_{HI}<0.5$ is $Def_{HI}$=0.06$\pm$0.28, consistent with the 
typical uncertainty in the estimate of $Def_{HI}$ for isolated field galaxies \citep{haynes,solanes96}.
\begin{figure*}
\centering
\includegraphics[width=14.cm]{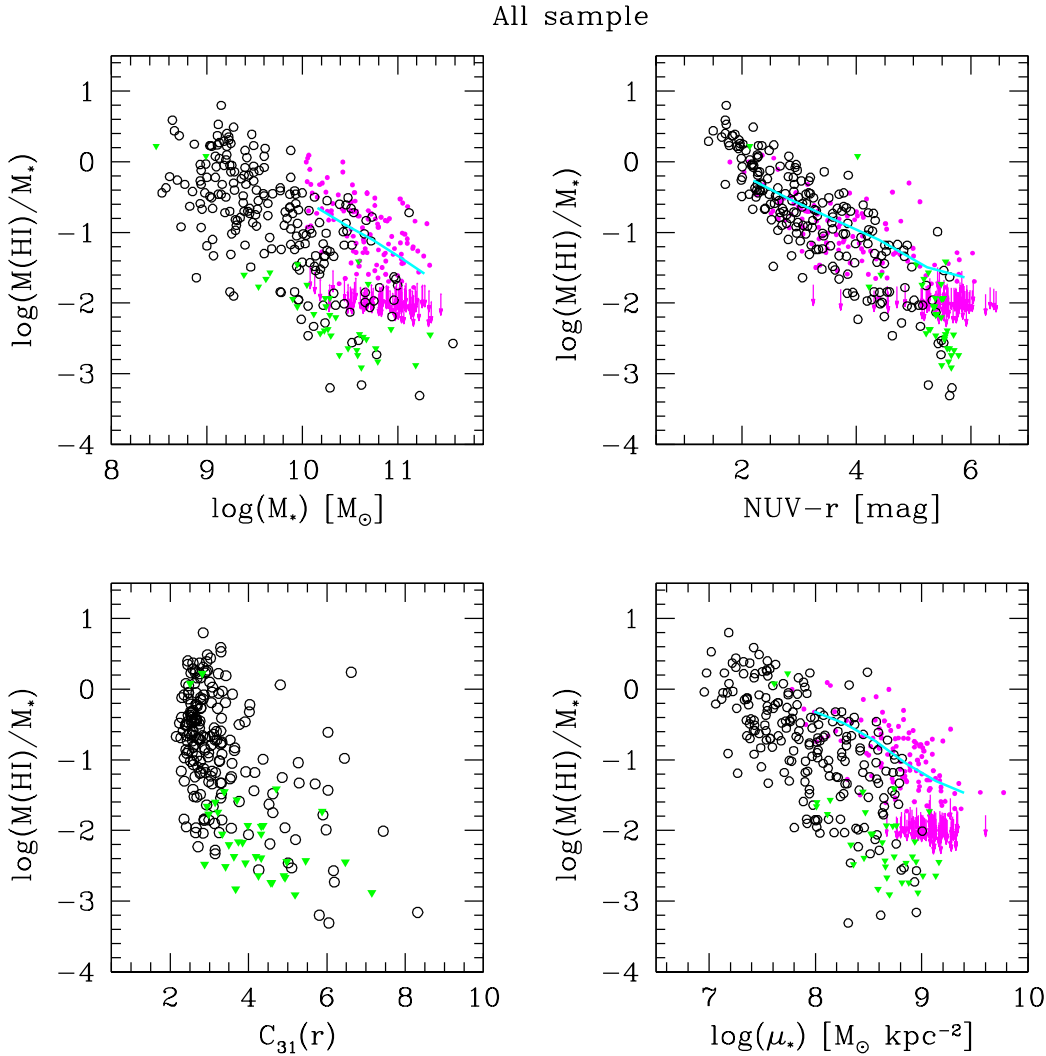}
\caption{The \hi~mass fraction of the HRS plotted as a function of stellar mass, $NUV-r$ colour, concentration index and 
stellar mass surface density. Black circles and green triangles represent detections and non-detections, respectively. GASS DR1 
detections and non-detections are indicated with magenta dots and arrows, respectively. The average scaling relations obtained 
from ALFALFA stacking of the GASS parent sample are indicated in cyan.}
\label{allscale}
\end{figure*}

Ultraviolet and optical broad-band photometry have been obtained from {\it Galaxy Evolution Explorer} (GALEX, \citealp{martin05}) and 
{\it Sloan Digital Sky Survey} DR7 (SDSS-DR7, \citealp{sdssDR7}) databases, respectively. 
GALEX near-ultraviolet (NUV; $\lambda$=2316 \AA: $\Delta \lambda$=1069 \AA) images have been mainly 
obtained as part of two on-going GALEX Guest Investigator programs (GI06-12, P.I. L. Cortese and 
GI06-01, GUViCS, \citealp{guvics}). 
Additional frames have been obtained from the GALEX GR6 public release. 
All frames have been reduced using the current version of the GALEX pipeline (ops-v7).  
The GALEX NUV and SDSS $g$,$r$,$i$ photometry for the HRS was determined as follows. 
The SDSS images were registered to the GALEX frames and convolved to the NUV 
resolution (5.3\arcsec, \citealp{morrissey07}). Isophotal ellipses were then fit to each 
image, keeping fixed the center, ellipticity and position angle (generally determined in the $i$-band). 
Asymptotic magnitudes have been determined from the growth curve obtained following 
the technique described by \cite{atlas2006} and corrected for Galactic extinction 
assuming a \cite{cardelli89} extinction law with $A(V)/E(B-V)=$3.1: i.e., 
$A(\lambda)/E(B-V)$=8.2 \citep{wyder07}, 3.793, 2.751 and 2.086 for $NUV$, $g$, $r$ and $i$, respectively.  
Stellar masses $M_{*}$ are determined from i-band luminosities $L_{i}$ using the $g-i$ colour-dependent 
stellar mass-to-light ratio relation from \cite{zibetti09}, assuming a \cite{chabrier} initial mass function (IMF):
\begin{equation}
\log(M_{*}/L_{i}) = -0.963 + 1.032*(g-i)
\end{equation}

The final sample used for the following analysis includes those HRS galaxies for which \hi, NUV and SDSS observations 
are currently available: 241 galaxies ($\sim$75\% of the whole HRS, namely 200 late- and 41 early-type galaxies). 

\section{The \hi~scaling relations of the HRS}
\label{secscale}
\begin{figure*}
\centering
\includegraphics[width=16.cm]{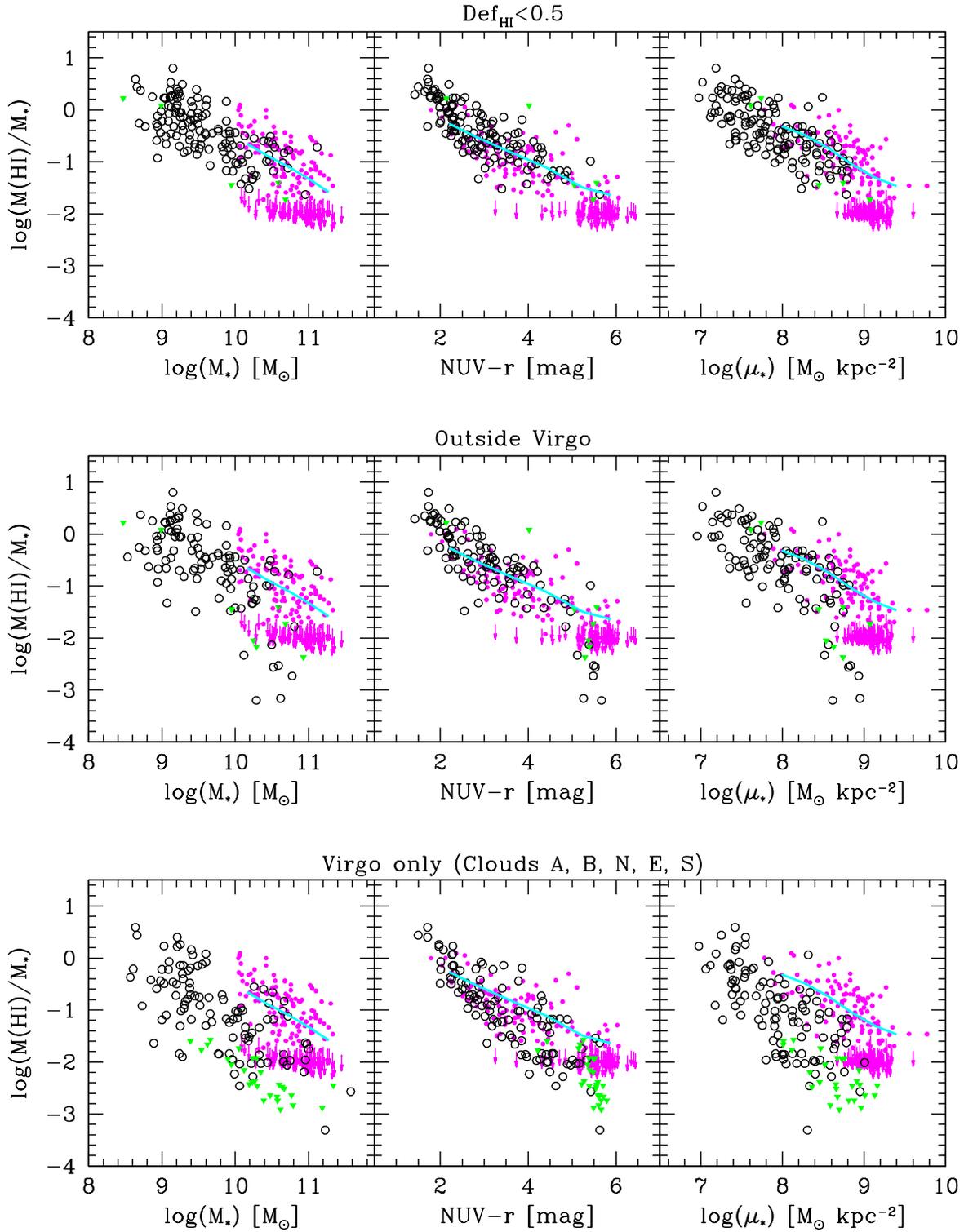}
\caption{The \hi~mass fraction as a function of stellar mass, $NUV-r$ colour and stellar mass surface density 
for \hi-normal galaxies (upper panel), galaxies outside the Virgo cluster (middle panel) and galaxies 
belonging to one of the Virgo cluster clouds (bottom panel). Symbols are as in Fig.~\ref{allscale}.}
\label{scaleenv}
\end{figure*}
\begin{figure*}
\centering
\includegraphics[width=17.5cm]{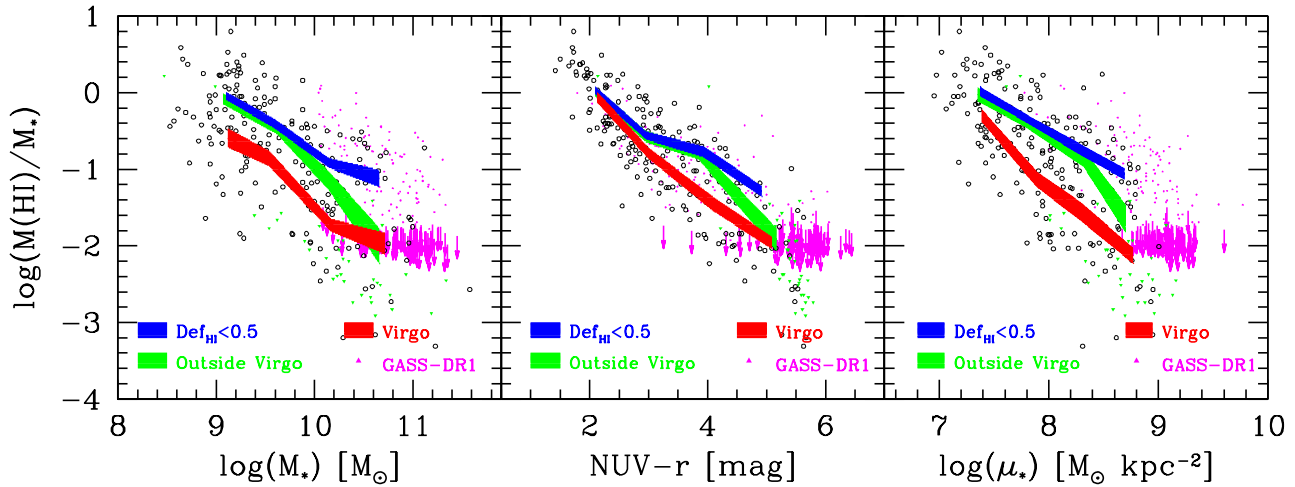}
\caption{The average \hi~mass fraction (i.e., $<log(M(HI)/M_{*})>$) scaling relations for different samples. 
\hi-normal, galaxies belonging to or outside Virgo are indicated by blue, red and green lines respectively. 
For comparison, GASS DR1 is shown in magenta.}
\label{scalemed}
\end{figure*}
In order to quantify how the \hi~gas fraction varies as a function of integrated galaxy properties, we plot in 
Fig.\ref{allscale} the \hi-to-stellar mass ratio as a function of stellar mass (upper-left panel), observed 
$NUV-r$ colour (upper-right), concentration index in r-band ($C_{31}(r)$, defined as the ratio between the radii 
containing 75\% and 25\% of the total $r$-band light) and stellar mass surface density (i.e., $M_{*}/(2\pi R_{50,i}^{2})$, 
where $R_{50,i}$ is the radius containing 50\% of the total $i$-band light i.e., the effective radius).   
For comparison, we also plot the data from the GASS data release 1 (\citealp{catinella10}, magenta symbols) and from the 
stacking of ALFALFA observations for the GASS parent sample (\citealp{fabello10}, cyan line). 
Over the range of stellar masses covered by the HRS, the \hi~gas fraction strongly anti-correlates with 
$M_{*}$, $NUV-r$ (a proxy for specific star formation rate, SSFR) and $\mu_{*}$, while a very weak non-linear trend is observed with the concentration index.
The tightest correlation is with the observed $NUV-r$ colour (Pearson correlation coefficient $r=-$0.89, dispersion 
along the y-axis $\sigma=$0.42 dex)\footnote{These parameters are obtained assuming the non-detections to their upper-limits.}, 
while the scatter gradually increases for the stellar surface density ($r=-$0.74, $\sigma=$0.61 dex) and mass ($r=-$0.72, $\sigma=$0.63 dex). 

Our findings extend very nicely the results of the GASS survey \citep{catinella10,fabello10} to lower stellar mass and surface densities, 
highlighting how the star formation rate, stellar mass and galaxy structure are tightly linked to the \hi~content of 
the local galaxy population. 
However, it is important to notice that the scaling relations for the HRS are slightly offset 
towards lower gas fractions when compared to the results of the GASS survey. 
This difference is easily understood when we consider that, by construction, nearly half of the galaxies 
in the HRS belong to the Virgo cluster and, as a whole, this sample might be biased towards gas-poor objects.
To test this and to characterize the \hi~scaling relations in different environments, we have divided 
our sample in three different subsets:  (a) galaxies with $Def_{HI}<0.5$ (i.e., \hi-normal, 135 galaxies), 
(b) galaxies outside the Virgo cluster (110 galaxies)  and (c) galaxies belonging to one of the Virgo cluster clouds 
(Virgo A, B, N, E and S as defined by \citealp{gav99}, 131 galaxies).
We note that, while (b) and (c) are complementary by construction, (a) may include both Virgo and field galaxies.   
The scaling relations for the three samples are plotted in Fig.~\ref{scaleenv}.
As expected, although for all samples the \hi~gas fraction decreases with stellar mass, colour and stellar mass 
surface density, galaxies in different environments show significantly different \hi~content.
Virgo galaxies have, on average, a lower \hi-to-stellar mass ratio than \hi-normal galaxies across the whole 
range of stellar masses and stellar mass surface densities covered by the HRS.
In fact, the dispersion in the two scaling relations drops from $\sim$0.62 dex for the whole sample to $\sim$0.37 dex 
for \hi-normal galaxies only.  

In order to quantify the difference between cluster and field, in Fig.~\ref{scalemed} we show 
the average trends (i.e., $<log(M(HI)/M_{*})>$) for each subsample.
The averages are measured by assuming the non detections to their upper-limit and by 
correcting for the fact that we do not have \hi~and ultraviolet data for all HRS galaxies.
In practice, we weighted each galaxy by the completeness of the sample in its bin of K-band luminosity.
The average scaling relations and the number of galaxies contributing to each bin 
are given in Table~\ref{scaletab}.
 
The difference between \hi-normal and cluster galaxies varies by $\sim$0.3 dex as a function of 
stellar mass (i.e., from $\sim$0.55 to 0.85 dex), while it increases from $\sim$0.25 to 1 dex when 
moving from disk- to bulge-dominated systems (i.e., from $\mu_{*}\sim$ 7.4 to 8.7 M$_{\odot}$ kpc$^{-2}$)\footnote{We 
remind the reader that, by construction, the HRS does not include intermediate-, low-mass 
early-type galaxies, typically found in high-density environments \citep{hughes09,cortese09}. Thus, for low stellar masses, low stellar mass surface densities 
and red colours, we might be underestimating the difference between cluster and field galaxies.}. 
Less strong is the environmental dependence of the gas fraction vs. $NUV-r$ relation. 
Blue-sequence galaxies ($NUV-r<$3.5 mag) have the same \hi~gas fraction regardless of the environment, 
consistent with a scenario in which actively star-forming cluster galaxies have just started their infall 
into the cluster center \citep{review,cortesecoma08}. On the contrary, red Virgo galaxies are significantly gas poorer than 
\hi-normal systems. It is important to note that this difference might be, at least partially, due to an 
extinction effect. While a significant fraction of red cluster galaxies are likely to be really passive systems, field red 
objects should be more heavily affected by internal dust attenuation. Therefore bins of observed $NUV-r$ colour might 
not directly correspond to bins of SSFR. We plan to investigate this issue in a future paper, 
where {\it Herschel} data will be used to properly correct for internal dust absorption.   
  
Interestingly, we find some differences also when comparing \hi-normal galaxies to objects outside the Virgo clusters. 
Since \hi-deficient galaxies are mainly/only found in clusters of galaxies \citep{haynes,ages1367,bosellicomaI}, it is common practice 
to assume that galaxies in the field, in pairs and loose groups are not \hi-deficient (and viceversa). 
Our analysis shows that a one-to-one correlation between \hi-normal galaxies and systems in low-/intermediate-density
environments apparently breaks at high stellar masses ($M_{*}>$10$^{10.4}$ M$_{\odot}$), stellar mass surface densities 
($\mu_{*}>$ 8.5 M$_{\odot}$ kpc$^{-2}$) and red colours ($NUV-r>$4.5 mag), where we find \hi-deficient galaxies also outside 
the cluster environment. It is important to note that a significant fraction of these 
gas-poor galaxies (i.e., 5 out of 7 and 6 out of 10 for 
the $\log(M(HI)/M_{*})$ vs. $\log(M_{*})$ and $\log(M(HI)/M_{*})$ vs. $\log(\mu_{*})$ relations, respectively) are early-type 
systems (S0a or earlier) for which the \hi~deficiency calibration is notoriously uncertain. 
Whether the typical red, massive, bulge-dominated galaxy contains a significant cold gas reservoir or not is still matter of debate.
These ranges of stellar masses, colour and stellar surface densities have not been adequately sampled by previous \hi~investigations 
and it is not completely surprising that we find a difference between \hi-normal and field galaxies.  

Whatever the origin of the difference between \hi-normal and `field' galaxies, the break observed in the \hi~scaling relations 
for galaxies outside the Virgo cluster is very similar to the transition observed in the colour-stellar mass and colour-stellar 
mass surface density diagrams (e.g., \citealp{kauffmann03b}). \cite{bothwell09} and \cite{catinella10}
already reported the presence 
of a transition in the \hi~gas fraction vs. stellar mass and stellar mass surface density relations, respectively (see also \citealp{kannappan04}). 
Our analysis confirms both results at once independently, suggesting that the typical transitions observed in the colour-mass 
diagram might be related to the transition in \hi~gas fraction showed in Figs.~\ref{scaleenv} and \ref{scalemed}.
Unfortunately, above the transition our statistics becomes too small and we will have to wait for larger \hi~surveys to 
characterize more in detail this break in the \hi~scaling relations. 
  
Finally, it is important to point out that, above the transition in stellar mass/surface density, the difference in \hi~gas fraction 
between cluster and field galaxies tends to disappear. This provides an indirect support to a scenario in which environmental 
effects are today mainly active on the population of low-/intermediate-mass disk galaxies, while the properties of high-mass/bulge-dominated 
systems are less affected by the local density. 

\begin{table}
\caption {The average scaling relations for the three samples discussed in Sec.~3.}
\[
\label{scaletab}
\begin{array}{cccc}
\hline\hline
\noalign{\smallskip}
\multicolumn{4}{c}{\rm HI-normal~(Def_{HI}<0.5)}\\
\noalign{\smallskip}
x  &  <x>  &  <\log(M(HI)/M_{*})>^{a} & N_{gal} \\
\noalign{\smallskip}
\hline
\log(M_{*})  &  9.10	     &	  -0.04\pm0.05  	    &	46   \\
     	     &  9.59	     &       -0.40\pm0.06	       &   44	\\
     	     &  10.15	     &       -0.92\pm0.06	       &   31	\\
     	     &  10.66	     &       -1.12\pm0.11	       &   10	\\
\noalign{\smallskip}
NUV-r  	     &    2.10  &    +0.03\pm  0.05    &       52   \\
  	     &    2.96  &    -0.56\pm  0.05    &       46   \\
  	     &    3.93  &    -0.77\pm  0.07    &       25   \\
  	     &    4.92  &    -1.29\pm  0.07    &	9   \\
\noalign{\smallskip}

\log(\mu_{*})  &     7.37   &	+0.02\pm	0.06	&	 36   \\
      	       &     7.80   &	-0.34\pm	0.05	&	 48   \\
      	       &     8.30   &	-0.75\pm	0.07	&	 32   \\
      	       &     8.69   &	-1.06\pm	0.07	&	 16   \\
\hline
\hline
\noalign{\smallskip}
\multicolumn{4}{c}{\rm Outside~Virgo}\\
\noalign{\smallskip}
x  &  <x>  &  <\log(M(HI)/M_{*})>^{a} & N_{gal} \\
\noalign{\smallskip}
\hline
\log(M_{*})      &   9.07   &    -0.08 \pm        0.07   &      36   \\
      		 &   9.62   &	 -0.47 \pm	  0.07   &	31   \\
      		 &  10.14   &	 -1.14 \pm	  0.12   &	29   \\
      		 &  10.66   &	 -2.00 \pm	  0.23   &	11   \\

\noalign{\smallskip}
NUV-r     	 & 2.10   &    +0.01 \pm	  0.06       &       36 \\
     	  	 & 2.95   &    -0.57 \pm       0.07	  &	  34 \\
     	  	 & 3.85   &    -0.79 \pm       0.08	  &	  20 \\
     	  	 & 5.17   &    -1.92 \pm       0.15	  &	  16 \\
\noalign{\smallskip}
\log(\mu_{*})      & 7.35  &     -0.02  \pm	   0.09       &       21 \\
     		   & 7.78  &	 -0.37  \pm	 0.07	    &	    40 \\
     		   & 8.32  &	 -0.86  \pm	 0.11	    &	    25 \\
     		   & 8.70  &	 -1.64  \pm	 0.18	    &	    21 \\
\hline
\hline
\noalign{\smallskip}
\multicolumn{4}{c}{\rm Virgo}\\
\noalign{\smallskip}
x  &  <x>  &  <\log(M(HI)/M_{*})>^{a} & N_{gal} \\
\noalign{\smallskip}
\hline
\log(M_{*})  &    9.12  &     -0.59  \pm      0.12	  &	    26  \\
     	     &    9.54  &     -0.86  \pm      0.10	  &	    33  \\
     	     &   10.17  &     -1.73  \pm      0.08	  &	    43  \\
     	     &   10.72  &     -1.97  \pm      0.14	  &	    22  \\
\noalign{\smallskip}
NUV-r       &	  2.13  &     -0.06  \pm       0.07	  &	     21 \\
      	    &	  3.00  &     -0.79  \pm       0.07	  &	     34 \\
      	    &	  4.09  &     -1.45  \pm       0.09	  &	     32 \\
      	    &	  5.09  &     -1.96  \pm       0.08	  &	     28 \\
\noalign{\smallskip}
\log(\mu_{*})   &	7.39 &      -0.30   \pm      0.08	 &	 29   \\
     	    	&   7.89 &	-1.13	\pm	 0.10	     &       35   \\
     	    	&   8.30 &	-1.56	\pm	 0.12	     &       34   \\
     	    	&   8.77 &	-2.13	\pm	 0.10	     &       30   \\
\hline
\hline

\end{array}
\]
a: We note that these are averages of $\log(M(HI)/M_{*})$ and not $M(HI)/M_{*}$ (as in \citealp{catinella10}). 
We preferred this approach because the distribution of \hi~gas fraction is closer to 
a log-normal than to a Gaussian.   
\end{table}

\begin{figure*}
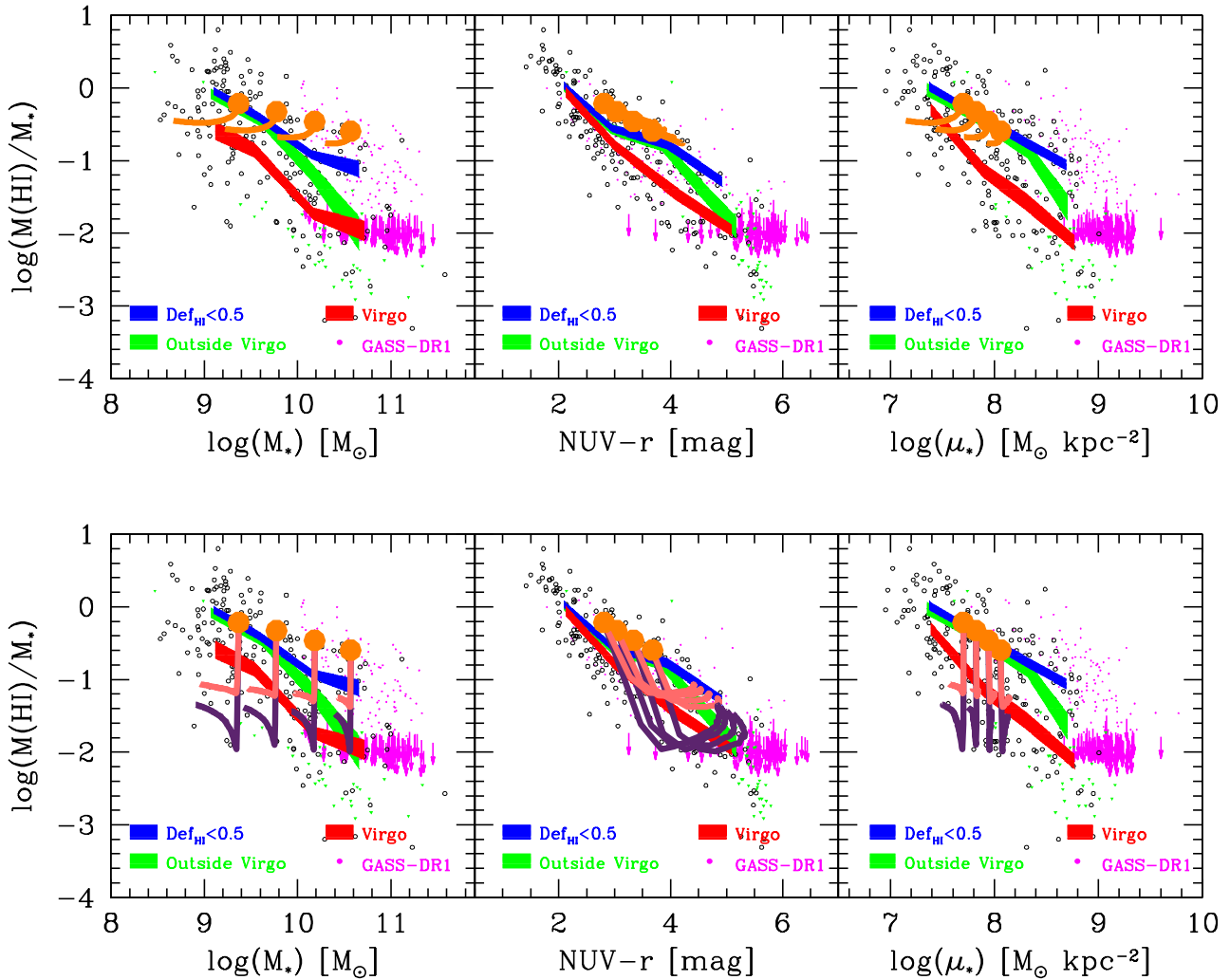

\centering
\includegraphics[width=17.5cm]{models_starvation.epsi}
\includegraphics[width=17.5cm]{models_rp.epsi}
\caption{Model predictions for a starvation (upper panels) and ram-pressure stripping (lower panels) event.
In case of starvation we considered ages between 0 and 9.5 Gyr, while for ram-pressure we restricted our analysis to events occurred 
between 0 and 5 Gyr ago. In both sets of panels, the unperturbed model is indicated by the filled circles. 
In the bottom panel, the two lines indicate models for different ram-pressure efficiencies:
$\epsilon _{0}=$0.4 (light curve) and 1.2 (dark curve) M$_{\odot}$ kpc$^{-2}$ yr$^{-1}$. Data are as in Fig.~\ref{scalemed}.
We note that these are not evolutionary tracks, but illustrate where a galaxy would lie today depending on the age of the interaction.}
\label{models}
\end{figure*}

\section{Comparison with models}
The results presented in \S~\ref{secscale} have shown that, at fixed stellar mass 
and surface densities, cluster galaxies have significantly lower \hi-content than 
galaxies in low density environments. On one side, these results confirm that 
\hi-deficient galaxies are mainly present in high density environments \citep{haynes}. 
On the other, they nicely complement previous analysis showing that cluster galaxies have in 
general a lower star formation rate than their field counterpart \citep{review,gav02,gomez03}, 
suggesting that the gas loss is behind the quenching of the star formation activity \citep{ha06,cortese09}.
Although this scenario is commonly accepted, there is still debate about the environmental 
mechanism(s) responsible for such differences between low- and 
high-density environments (e.g., \citealp{weinmann11}). 
In particular, it is not clear whether the \hi~is stripped directly from the disk 
via a strong interaction with the dense intra-cluster medium (i.e., ram-pressure 
stripping; \citealp{GUNG72}), or if the gas is just removed from the extended gaseous halo surrounding 
the galaxy, preventing further infall (i.e., starvation; \citealp{larson80}). 
\cite{dEale} have recently used multi-zone chemical and spectrophotometric models to constrain 
the evolutionary history of dwarf galaxies in the Virgo cluster and to discriminate between 
the ram-pressure and starvation scenarios. 
They showed that the properties of most of the dwarfs dominating the faint end of the Virgo luminosity function 
are consistent with a scenario in which these galaxies were initially star-forming systems, accreted by the cluster 
and stripped of their gas by one or more ram pressure stripping events.  

Here, we use the same approach adopted by \cite{dEale} to carry out a similar test for intermediate and massive galaxies in the Virgo cluster. 
The details of the models here adopted can be found in \citet[Appendix B]{dEale}.
Briefly, the evolution of galaxies is traced using the multi-zone chemical and spectrophotometric model of 
\cite{boissier00}, updated with an empirically determined star formation law \citep{boissier03} relating the 
star formation rate to the total gas surface densities. 
Following \cite{dEale}, we modified the original models to simulate the effects induced by the interaction with the cluster environment. 
In the starvation scenario we simply stopped the infall of pristine gas in the model. 
We investigate interactions that started between 0 and 9.5 Gyr ago. 
The ram pressure stripping event is simulated by assuming a gas-loss rate inversely proportional to the potential of 
the galaxy, with an efficiency depending on the density of the intra-cluster medium (taken from \citealp{vollmer01}). 
We assume two different values for the peak efficiency of ram-pressure stripping 
($\epsilon _{0}=$0.4 and 1.2 M$_{\odot}$ kpc$^{-2}$ yr$^{-1}$) and an age of the interaction between 0 and 5 Gyr ago.
Models have been generated assuming a spin parameter $\lambda=$0.05 and four different rotational velocities 
(V$_{C}=$100, 130, 170, 220 km s$^{-1}$), in order to cover the whole range of stellar masses spanned by the HRS. 
In order to convert the output of the model into the observables here considered, we proceeded as follows. 
Stellar masses have been converted from a \cite{kroupa93}, used in the model, to a \cite{chabrier} IMF by adding 
0.06 dex \citep{bell03,gallazzi08}. Stellar mass surface densities have been determined using the effective 
radius in i-band, as done for our data. \hi~masses are obtained from the total gas mass by removing the contribution of helium and heavy 
elements (i.e., a factor 1.36) and assuming a molecular-to-atomic hydrogen 
gas ratio of 0.38, independent of stellar mass \citep{saintonge11}. $NUV-r$ colours have been `reddened' by assuming 
an average internal extinction $A(NUV)=$1 mag. This value is quite arbitrary and will likely depend on the 
evolutionary history of each galaxy. However, this assumption does not affect our main conclusions.   
Finally, we note that the goal of this exercise is to establish which scenario is more consistent with out data, not to 
determine the best model fitting our observations. There are many free parameters and assumptions in the 
modeling that could be tweaked in order to better match our observations, but looking for an exact fit does not 
seem meaningful given the simple description adopted here. 

In Fig.~\ref{models} (top panels) we compare the predictions of the starvation model with our data. 
The filled circles indicate the predictions for the unperturbed models. 
The models seem to fairly reproduce the $NUV-r$ and $\mu_{*}$ vs. gas fraction scaling relations 
observed for \hi-normal and field galaxies. 
However, a significant difference between data and models is observed in the $M(HI)/M_{*}$ vs. $M_{*}$ relation: 
for very high stellar masses (rotation velocities), the unperturbed model predicts gas fractions $\sim$0.5 dex higher 
than the average observed value. 
This likely reflects the fact that the model is only valid for disk galaxies and does not include any bulge component. 
It is thus not surprising that at high masses, where the fraction of bulge-dominated galaxies increases, 
the predictions we obtain are not representative of the whole population (although still 
well within the range of gas fractions covered by the observations). This is also confirmed by the fact that 
the model covers just the range of stellar mass surface densities typical of disk galaxies.
    
By comparing the predictions of the starvation model with our data, it clearly emerges that starvation is only able 
to mildly affect the \hi~content, even for very old events. 
The \hi~gas fraction decreases by only \ls0.25 dex, i.e. within the intrinsic dispersion of the scaling relation for \hi-normal galaxies. 
This decrease is even lower in the $M(HI)/M_{*}$ vs. $NUV-r$ plot, where galaxies affected by starvation lie on the main relation observed 
for unperturbed galaxies. It is thus impossible to reproduce the difference between cluster and field galaxies 
by just stopping the infall of gas onto the disk.
 
Much more promising are the results obtained for the ram-pressure stripping model (Fig.~\ref{models}, bottom panels).
For all the three scaling relations here observed, ram-pressure stripping is able to fairly reproduce the difference 
observed between cluster and field galaxies. A maximum efficiency of 0.4  M$_{\odot}$ kpc$^{-2}$ yr$^{-1}$ (light curve) is sufficient 
to reproduce the average decrease in \hi~content for Virgo galaxies, while a higher efficiency (1.2 
M$_{\odot}$ kpc$^{-2}$ yr$^{-1}$, dark curve) is necessary to explain some of the most \hi-deficient systems in our sample. 
As expected, the $M(HI)/M_{*}$ vs. $NUV-r$ is the relation least affected by the environment. Once the gas is removed, 
the star formation is quickly reduced and the perturbed galaxy will not shift considerably from the main relation.
In conclusion, only the direct removal of gas from the star-forming disk is able to reproduce the significant difference 
between the \hi~scaling relations of cluster and field galaxies.
\begin{figure}
\centering
\includegraphics[width=8.5cm]{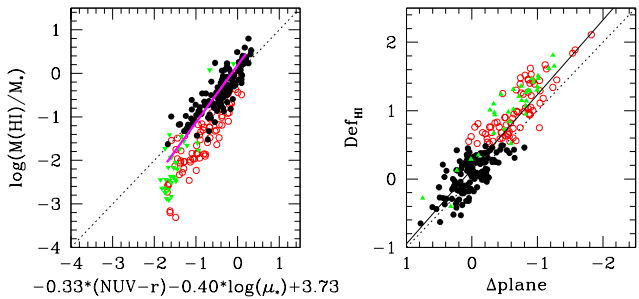}
\caption{Left panel: The \hi~gas fraction plane for \hi-normal ($Def_{HI}<$0.5) HRS galaxies.The \hi~gas fraction plane obtained for GASS galaxies is indicated 
by the magenta line. Right panel: The correlation between \hi~deficiency and distance (along the y-axis) from the plane. The solid 
line indicates the best linear fit to the \hi~detections. 
In both panels, \hi-normal, \hi-deficient galaxies and non-detections are indicated 
with black filled, red empty circles and green triangles, respectively, and the dotted line shows 
the 1:1 anti correlation.}
\label{plane}
\end{figure}

\section{The \hi~gas fraction plane and the \hi~deficiency parameter}
In the previous sections we have confirmed the existence of a tight relation between 
\hi~gas fraction, stellar mass surface density and $NUV-r$ colour.
Recent works have used these three quantities to define a `gas fraction plane' to be 
used in order to determine gas fractions for large samples lacking \hi~observations 
(\citealp{cheng_gfrac09}, using optical colours) and to isolate galaxies in 
the process of accreting or loosing a significant amount 
of gas \citep{catinella10}.
As discussed by \cite{cheng_gfrac09}, the existence of this plane can be seen as a direct 
consequence of the Kennicutt-Schmidt relation.  
In order to test how good the gas fraction plane is as a proxy for \hi~deficiency, we 
decided to compare the two approaches for our sample. 
We thus fitted a plane to the two-dimensional relation between \hi~mass fraction, 
stellar mass surface density and $NUV-r$ colour for \hi-normal galaxies only ($Def_{HI}<$0.5), 
following the methodology outlined in \cite{bernardi03}, i.e., by minimizing the 
residuals from the plane. 
The best fit to the plane is 
\begin{equation}
\log \Big(\frac{M(HI)}{M_{*}}\Big) =  -0.33(NUV-r) -0.40\log(\mu_{*}) + 3.73 \newline
\end{equation} 
and has a dispersion of only $\sim$0.27 dex, comparable to the typical uncertainty in the 
\hi~deficiency parameter.
The plane so obtained is illustrated in Fig.~\ref{plane} (left panel) and is fairly consistent with 
the one determined for the GASS sample by \citet[magenta line]{catinella10}. Given the different 
datasets and selection criteria adopted by the two surveys, this agreement is quite remarkable 
and indicates that both \hi-normal HRS and GASS samples provide a fair representation 
of the \hi~properties of massive galaxies in the local 
universe. 
As expected, \hi-deficient galaxies (empty circles) are significantly offset from the plane, having 
a lower \hi-content than what expected for their colour and stellar surface density. 
The distance from the plane along the y-axis (i.e., $log(M_{HI}/M_{*})_{obs} - log(M_{HI}/M_{*})_{plane}$) strongly 
anti correlates with the \hi~deficiency (Fig.~\ref{plane}, right panel). The best linear fit for the \hi~detections is
\begin{equation}
Def_{HI} = (-1.09\pm 0.04) \times \Delta plane + (0.14\pm 0.02)
\end{equation}
with $\sim$0.25 dex dispersion. Again, the scatter is similar to the intrinsic scatter in the definition of the \hi~deficiency, 
suggesting that the \hi~plane is a tool as good as the \hi~deficiency to isolate extremely \hi-rich 
or \hi-poor galaxies, in particular when accurate morphological classification is not available.
\begin{figure}
\centering
\includegraphics[width=8.5cm]{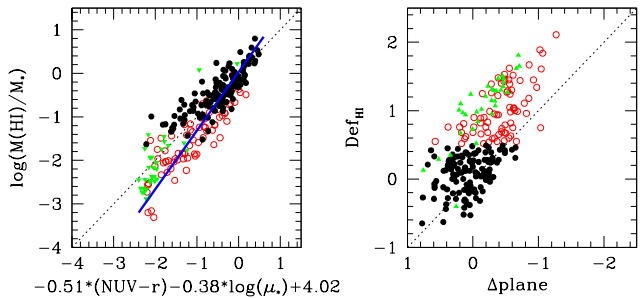}
\caption{Left panel: The \hi~gas fraction plane for HRS galaxies outside the Virgo cluster.
The \hi~gas fraction plane obtained for \hi-normal HRS galaxies (Fig.~4) is indicated 
by the solid line. Right panel: The correlation between \hi~deficiency and distance (along the y-axis) from the plane. 
Symbols are as in Fig.~\ref{plane}.}
\label{planefield}
\end{figure}

Although the results obtained here are very promising, we want to conclude this section with some notes of caution 
in the definition, and use, of the \hi~gas fraction plane.  
The definition of a plane for \hi-normal galaxies is justified by the fact that, for this sample, 
we observe a linear correlation between $\log(M(HI)/M_{*})$ and both $NUV-r$ colour and $\log(\mu_{*})$. 
As discussed in the previous section, these linear relations are not valid for typical `field' galaxies across 
the whole range of parameter here investigated, but we find a break for massive, bulge-dominated, red galaxies. 
Therefore, when these objects are included, the concept of a plane is no longer justified. 
In order to show how important is the sample selection, we plot in Fig.~\ref{planefield} the plane obtained 
for galaxies outside the Virgo cluster and compare it with what previously obtained for \hi-normal galaxies 
(solid line). The two planes are significantly different, in particular at low \hi~gas fractions where 
there is an offset up to a factor $\sim$4 in the estimate of the \hi~mass. 
Incidentally, this shows once more how `field' galaxies is not the same as `\hi-normal' systems.
Moreover, this new plane is almost useless to isolate \hi-deficient galaxies. 
\hi-poor systems are no longer strong outliers and the relation between \hi~deficiency and 
distance from the plane is too scattered ($\sigma\sim$0.4 dex) to be useful. 
We note that the good agreement between our plane for \hi-normal galaxies and the one obtained 
using GASS detections is due to the fact that both datasets sample the stellar mass 
surface density regime below the break in \hi~gas fraction. 
Notice that, if the GASS detection limit was much deeper than 1.5\% gas fraction (which roughly 
corresponds to the limit separating \hi-normal and \hi-deficient galaxies at high stellar masses, 
see Fig.~\ref{scaleenv}), then this survey would include detections above the $\mu_{*}$ break, where 
the definition of a plane is no longer justified. 

In conclusion, the \hi~gas fraction plane is a promising tool to estimate gas fractions and isolate outliers, 
but only if calibrated on well defined samples of `normal' unperturbed galaxies.

\section{Conclusions}
In this paper we have shown the relations between \hi-to-stellar mass ratio, 
galaxy structure and colour for the HRS, a volume-, magnitude-limited sample of nearby galaxies. 
\hi~gas fractions anti correlate with stellar mass, stellar mass surface 
density and $NUV-r$ colour, confirming 
previous results (e.g., \citealp{catinella10}).
We show that these trends are observed in both low- and high-density environments, 
with the only difference that the scaling relations for cluster galaxies are shifted towards lower gas fractions.
Thus, the cluster environment seems to have just a secondary (although important) 
role on galaxy evolution, with the efficiency of gas consumption and star formation mainly 
driven by the intrinsic `size' (e.g., total mass) of the system \citep{phenomen,boselli01,kauffmann03b}. 
Our findings nicely complement the scaling relations between SSFR 
and galaxy properties revealed by large optical and ultraviolet surveys \citep{blanton05,schiminovich07}. 
Naturally, because the \hi~is the fuel for star formation, it is likely that the 
link between SSFR and integrated properties is driven by the scaling relations 
presented here. This is also supported by the presence of a break in 
the $M(HI)/M_{*}$ vs. $\mu_{*}$ and $M(HI)/M_{*}$ vs. $M_{*}$ relations, 
very close to the typical value characterizing the transition between the blue cloud and red sequence.   

In order to shed light on the environmental mechanism responsible for the lower \hi~content 
of cluster galaxies, we compared our results with the predictions of the chemo-spectrophotometric 
model by \cite{boissier00}. We find that simply halting the infall of pristine gas from the halo, 
mimicking the effect of starvation, is not sufficient to reproduce our observations.
Only the stripping of gas directly from the star-forming disk (i.e., ram-pressure) is able 
to match our data. This naturally explains why \hi-deficient galaxies 
are mainly/only found in very high-density environments, where ram-pressure stripping 
is efficient. However, it is important to note that our results do not imply that starvation 
is not playing any role on galaxy evolution. They just show that, if galaxies suffer of starvation, 
their star formation activity and gas content can be changed, but the shape of the main 
\hi~scaling relations will not be significantly altered.  
We remind the reader that these results are only valid for disk galaxies. 
At high masses, where bulge-dominated systems 
dominate, not only our model cannot be used but, most importantly, the difference between field and cluster 
starts to disappear, suggesting that the origin of the lower \hi~content of these objects cannot only be 
due to the cluster environment. 

One of the interesting outcomes of our analysis is that, at high stellar masses, `field' galaxies 
are not the same as `\hi-normal' objects. If confirmed, this clear difference 
might have important implications in our understanding of environmental effects in the local universe. 
In particular, one of the main issues for environmental studies is to reconcile the evidence 
that the effects of the environment start at densities 
typical of poor groups (\citealp{blanton09} and references therein) with the fact that \hi-deficient objects are 
only observed in clusters of galaxies. Our analysis suggests that, from a statistical point of view, 
\hi-deficient galaxies are clearly segregated in high-density environments only for 
$M_{*}$\ls10$^{10}$ M$_{\odot}$ and $\mu_{*}$\ls10$^{8}$ M$_{\odot}$ kpc$^{-2}$ or, 
in other words, that intermediate-/low-mass gas-poor quiescent disks are only found in clusters of galaxies. 
Thus, the results obtained from \hi~data turn out to be fully consistent with what shown by optical studies: 
i.e., the intermediate-/low-mass part of the red sequence is only populated in clusters of 
galaxies \citep{baldry06,haines08,gavazzi2010}.

Finally, we have confirmed that the \hi~gas fraction plane defined by \cite{catinella10} can be used as an alternative 
to the \hi~deficiency parameter. However, we have also illustrated how important is the sample selection in 
the definition of the \hi~gas fraction plane and that, as for the \hi~deficiency, its use is mainly 
valid for galaxies below the threshold in $\mu_{*}$. 

Our results clearly highlight the importance of \hi~scaling relations for environmental studies and 
the necessity to extend this approach to larger samples and to a wider range of environments in order to 
understand in detail the role played by nurture in galaxy evolution.

\section*{Acknowledgments}
We thank the GALEX Time Allocation Committee for the time granted to the HRS and GUViCS projects. 

GALEX is a NASA Small Explorer, launched in 2003 April. 
We gratefully acknowledge NASA's support for construction, operation and science analysis 
for the GALEX mission, developed in cooperation with the Centre National d'Etudes Spatiales (CNES) 
of France and the Korean Ministry of Science and Technology.

This publication makes use of data products from Two Micron All Sky Survey, 
which is a joint project of the University of Massachusetts and the Infrared Processing and Analysis 
Center/California Institute of Technology, funded by the National Aeronautics and Space 
Administration and the National Science Foundation.

Funding for the SDSS and SDSS-II has been provided by the Alfred P. Sloan Foundation, the Participating Institutions, 
the National Science Foundation, the U.S. Department of Energy, the National Aeronautics and Space Administration, 
the Japanese Monbukagakusho, the Max Planck Society, and the Higher Education Funding Council for England. 
The SDSS Web Site is http://www.sdss.org/.

The SDSS is managed by the Astrophysical Research Consortium for the Participating Institutions. 
The Participating Institutions are the American Museum of Natural History, Astrophysical Institute Potsdam, 
University of Basel, University of Cambridge, Case Western Reserve University, University of Chicago, 
Drexel University, Fermilab, the Institute for Advanced Study, the Japan Participation Group, Johns Hopkins University, 
the Joint Institute for Nuclear Astrophysics, the Kavli Institute for Particle Astrophysics and Cosmology, 
the Korean Scientist Group, the Chinese Academy of Sciences (LAMOST), Los Alamos National Laboratory, 
the Max-Planck-Institute for Astronomy (MPIA), the Max-Planck-Institute for Astrophysics (MPA), 
New Mexico State University, Ohio State University, University of Pittsburgh, University of Portsmouth, 
Princeton University, the United States Naval Observatory, and the University of Washington.

We acknowledge the use of the NASA/IPAC Extragalactic Database (NED) which is operated by the 
Jet Propulsion Laboratory, California Institute of Technology, under contract with the National 
Aeronautics and Space Administration and of the GOLDMine data base.

\end{document}